\documentclass[11pt,reqno]{amsart}
\usepackage{amsmath,amssymb,mathrsfs,graphicx}
\usepackage[utf8x]{inputenc}
\usepackage[T1]{fontenc}
\usepackage[USenglish,english]{babel}
\usepackage{amsmath}
\usepackage{amsfonts}
\usepackage{latexsym}
\usepackage{cite}
\usepackage{float}
\usepackage{tikz}
\usepackage{pgfplots}
\usepackage{caption}
\usepackage{booktabs} 
\usepackage{subcaption}
\usepackage{mathtools}      
\mathtoolsset{showonlyrefs} 
\usepackage[export]{adjustbox}
\usepackage{amsthm}
\usepackage{amssymb,amscd}
\usepackage{mathtools}
\usepackage{xargs}
\usepackage{graphicx}
\usepackage{color}
\usepackage{enumitem}
\usepackage{multirow}
\usepackage{lmodern}

\usepackage[text={6.5in, 8in}, centering]{geometry}
 
\usepackage{parskip}
\usepackage{microtype}

\usepackage{hyperref}

\usepackage{subcaption}

\newtheorem*{thm*}{Theorem}

\usepackage{appendix}

\theoremstyle{definition}

\title[]{Understanding Surgical smoke in Laparoscopy through Lagrangian Coherent Structures}
\author[]{Sandeep Kumar}
\address[]{Department of Quantitative methods, CUNEF Universidad, Madrid, Spain.}
\email{sandeep.kumar@cunef.edu}
\author[]{Caroline Crowley}
\address[]{School of Mechanical and Material Engineering, University College Dublin, Dublin, Ireland}
\email{caroline.crowley@ucdconnect.ie}
\author[]{Mohammad Faraz Khan}
\address[]{UCD Centre of Precision Surgery, School of Medicine, University College Dublin, Dublin, Ireland}
\email{mohammad.khan@ucd.ie}
\author[]{Miguel D. Bustamante}
\address[]{School of Mathematics and Statistics, University College Dublin, Dublin, Ireland}
\email{miguel.bustamante@ucd.ie}
\author[]{Ronan Cahill}
\address[]{Department of Surgery, Mater Misericordiae University Hospital, Dublin, Ireland}
\email{ronan.cahill@ucd.ie}
\author[]{Kevin Nolan}
\address[]{School of Mechanical and Material Engineering, University College Dublin, Dublin, Ireland}
\email{kevin.nolan@ucd.ie}

\date{\today}	
\begin{document}
\newenvironment{red}{\textcolor{red}}

\maketitle
\begin{abstract}
In laparoscopic surgery, one of the main byproducts is the gaseous particles, called surgical smoke, which is found hazardous for both the patient and the operating room staff due to their chemical composition, and this implies a need for its effective elimination. The dynamics of surgical smoke are monitored by the underlying flow inside the abdomen and the hidden Lagrangian Coherent Structures (LCSs) present therein. In this article, for an insufflated abdomen domain, we analyse the velocity field, obtained from a computational fluid dynamics model, {first, by calculating the flow rates for the outlets} and then by identifying the patterns which are responsible for the transportation, mixing and accumulation of the material particles in the flow. From the finite time Lyapunov exponent (FTLE) field calculated for different planar cross sections of the domain, we show that these material curves are dependent on the angle, positions and number of the outlets, and the inlet. The ridges of the backward FTLE field reveal the regions of vortex formation, and the maximum accumulation, details which can inform the effective placement of the instruments for efficient removal of the surgical smoke.
	 
\end{abstract}
\section{Introduction}

Minimally Invasive Surgery (MIS), or `keyhole' (laparoscopic) surgery has become increasingly popular, thanks to its effectiveness and economical advantages \cite{sauerland2006laparoscopy}. A typical laparoscopic operation involves piercing the human abdomen wall in four or five places through which, surgical instruments including a camera are passed inside the body. The hole thus created, is held open through an inlet port, i.e., the trocar, via which carbon dioxide gas is passed inside the abdomen to maintain a working space (pneumoperitoneum). In this way, with an insufflated abdomen, with more space, the surgeons can operate while viewing the telescopic image via an attached camera and light source connected to a video monitor. Some of the main benefits of laparoscopic surgery compared to conventional surgery are faster recovery and shorter stay at the hospital, reduced scarring due to small wounds, less pain and bleeding post-surgery, thus, a shorter period of short-term disability, etc. 

Laparoscopic surgery employs long slender surgical instruments that are passed through the valved trocars. These instruments cut the tissues inside the abdomen using a high-frequency alternating electrical current \cite{McCauley2010}. The gaseous byproduct of this process is called `surgical smoke' and it not only obscures the surgeon's vision through the camera, it can be observed that smoke and gas escape through the trocar \cite{FCC}. Several studies have addressed the hazardous nature of surgical smoke to the long-term health of surgical teams \cite{hardy2021aerosols,mac2022aerosol}. For instance, the gas leaks have been quantified and well-studied with high-speed Schlieren optical imaging and the latest image processing techniques \cite{CDKFN}. {Among other relevant concerns, it is worth mentioning  the surgical site infection, which can occur post surgery at the incision port, and is affected by particle distribution inside the the operating room (OR). The later has been well studied highlighting the effects and prevension in \cite{tan2023numerical,wong2022effects,liu2021prevention}. } Thus, to ensure the safety of the operating room (OR) staff and environment, a protected and safe MIS is desired and to that end, understanding the dynamics of gas both inside and outside the abdomen becomes very important \cite{crowley2022cfd}. It is also worth pointing out that surgical masks are not traditionally considered Personal Protective Equipment (PPE) but part of the infection control chain for the patient. PPE is the last resort in the hierarchy of controls and this work is motivated by a need for better engineering controls. 

With this aim, one possible direction is to model the underlying fluid phenomenon mathematically and study the velocity field. To begin with, a natural choice is to employ the Computational fluid dynamics (CFD) based techniques that can describe flow dynamics and the transport of smoke particles as reliable experimental data are sometimes difficult to obtain and can be expensive and time-consuming \cite{CY}. Recently, CFD modelling has been employed to understand the behaviour of surgical smoke as it is released outside the human body \cite{NSCTC,crowley2022cfd}. For instance, in \cite{NSCTC}, through a numerical model and simulations, smoke generation, evacuation and quantification of its composition have been described as well as compared with experimental data, while \cite{crowley2022cfd} provides novel information about the flow dynamics of gas leaks from trocars during laparoscopic surgery by studying the trajectories of particles ejected in the leaks. On the other hand, equally important and not much addressed is the action of gas as it enters the abdomen, which also influences the movement of smoke particles. In this article, we concentrate on this scenario and analyse different setups of laparoscopic surgery involving a change in positions, angles and number of outlets and the inlet. The latter is important because the configuration of trocars incisions varies from one procedure to another and the angle of the trocars changes throughout an operation as instruments are manipulated.  This is essentially an impinging jet problem, which results in the formation of vortical flow structures that will be constrained by the shape of the abdomen and the angle of the inflow jet. Surgical smoke is typically allowed to diffuse within the pneumoperitoneum and then vented into the operating theatre airspace in the breathing zones of theatre staff. Thus, a means to eliminate it at source requires an understanding of the flow dynamics inside the abdomen, which, in turn, would also complement the results in \cite{crowley2022cfd} and provide a complete picture of the surgical smoke problem.

Note that from the velocity field calculated from the CFD simulations one can study the particle trajectories; however, the instantaneous velocity field does not always reveal the behaviour of actual trajectories as the instantaneous streamlines can diverge from actual particle trajectories very quickly. Therefore, in a Lagrangian framework, the finite time Lyapunov exponent (FTLE) is defined which is calculated from the particle trajectories and hence, can explain the integrated effect of the flow \cite{shadden2005definition}. In other words, it measures the amount by which particles separate in a given time interval or a finite time average of the maximum expansion rate for a pair of particles advected in the flow.  Indeed, from the FTLE field, which is both space and time-dependent, complex flow patterns such as vortex formation and its development can be detected; moreover, the ridges of the field form separatrices in time-dependent systems, which are called Lagrangian Coherent Structures (LCS)  \cite{haller2015lagrangian}. These material curves can be identified in both forward and backward FTLE fields and are analogue of stable and unstable manifolds in time-independent systems, respectively, and divide dynamically distinct regions in the flow and reveal geometry which is often hidden when viewing the vector field or even trajectories of the system. Hence, they often provide an effective tool in analyzing systems with general time-dependence, especially for understanding transport of particles \cite{beron2013objective}. For instance, their applicability has been demonstrated in the blood arteries flow, identifying regions of circulation, and predicting sources and fate of debris in the oceanic flow \cite{vetel2009lagrangian, suara2020material,fluids1040038}. 

In this work, we perform an in-depth analysis of the velocity field within the insufflated abdomen domain, where we identify these LCS patterns that govern the transportation, mixing, and accumulation of material particles within the flow. {Note that,  this problem has gained much interest only recently, and the most of the work is performed by clinicians with limited engineering rigour. Although there are some studies involving CFD, none has considered the internal flow, nor has shown the smoke concentration the way LCS can; moreover, only simpler geometries and single arrangements have been analysed \cite{wang2022simulation, NSCTC}. This article advances the subject by moving to a more realistic 3D arrangement of ports and using LCS to give a more rigorous understanding of the transport of smoke particles. More preciesely}, we calculate the backward FTLE field, investigating various planar cross-sections of the domain. Remarkably, our findings demonstrate that these material curves are influenced by factors such as the angle, positions, and the number of outlets and the inlet, adding further complexity to the smoke's behaviour. The ridges of the FTLE field serve as vital indicators, revealing regions of vortex formation and maximum accumulation. Such intricate details provide invaluable insights for strategically placing instruments, which can help with the efficient and timely removal of surgical smoke. {In the preliminary version of this article, we have performed a similar analysis on velocity field data obtained through CFD simulations on a two dimensional domain, while this article takes a more realistic approach by working with the planar cross-section of a three dimensional domain \cite{kumar2022surgicalsmoke}. Hence, we offer a step forward in guiding the design of effective smoke evacuation technologies.} The structure of the article is the following. In Section \ref{sec:ProbSetup}, we briefly describe the problem definition, experimental setup and different types of laparoscopic surgery. We elaborate on the CFD model, mesh generation, and the relevant system of equations supplemented with suitable boundary conditions and parameters that are essential for the validity of the model.  In Section \ref{sec:ResultsDiscussion}, with a brief introduction to LCS, we consider different geometrical scenarios and analyse the velocity field obtained from the CFD model. 
Finally, the results are summarized in Section \ref{sec:Conclusions}.

\section{Problem definition, Experimental set-up and a CFD model}
\label{sec:ProbSetup}
\begin{figure}
	\centering
	\begin{subfigure}[b]{0.2\textwidth}
		\centering
		\includegraphics[width=\textwidth]{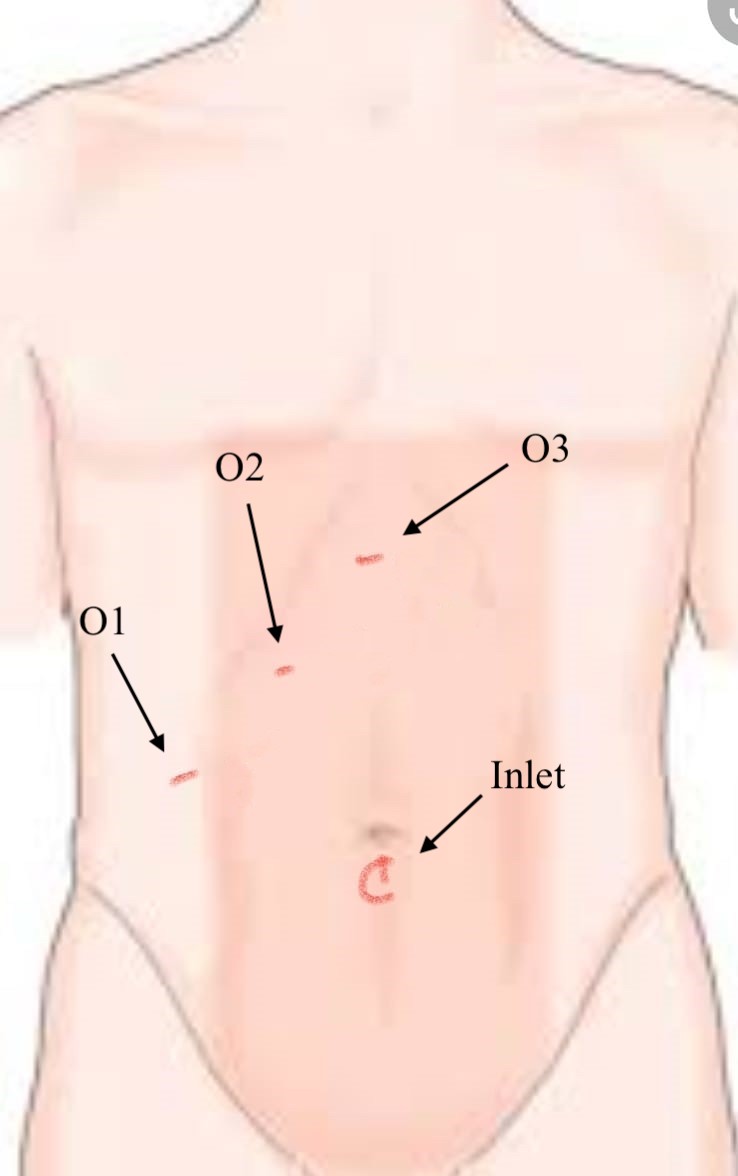}
		\caption{Case 1}
		\label{fig:Case1}
	\end{subfigure}
	\hfill
	\begin{subfigure}[b]{0.2\textwidth}
		\centering
		\includegraphics[width=\textwidth]{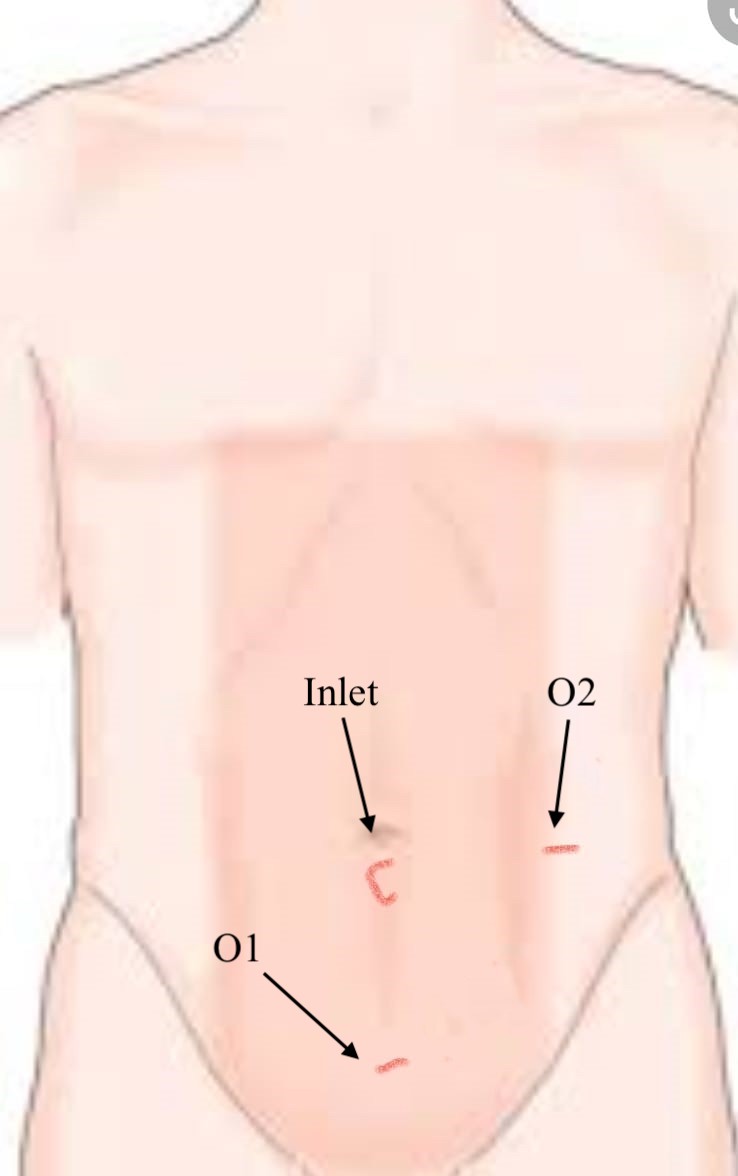}
		\caption{Case 2}
		\label{fig:Case2}
	\end{subfigure}
	\hfill		
	\begin{subfigure}[b]{0.2\textwidth}
		\centering
		\includegraphics[width=\textwidth]{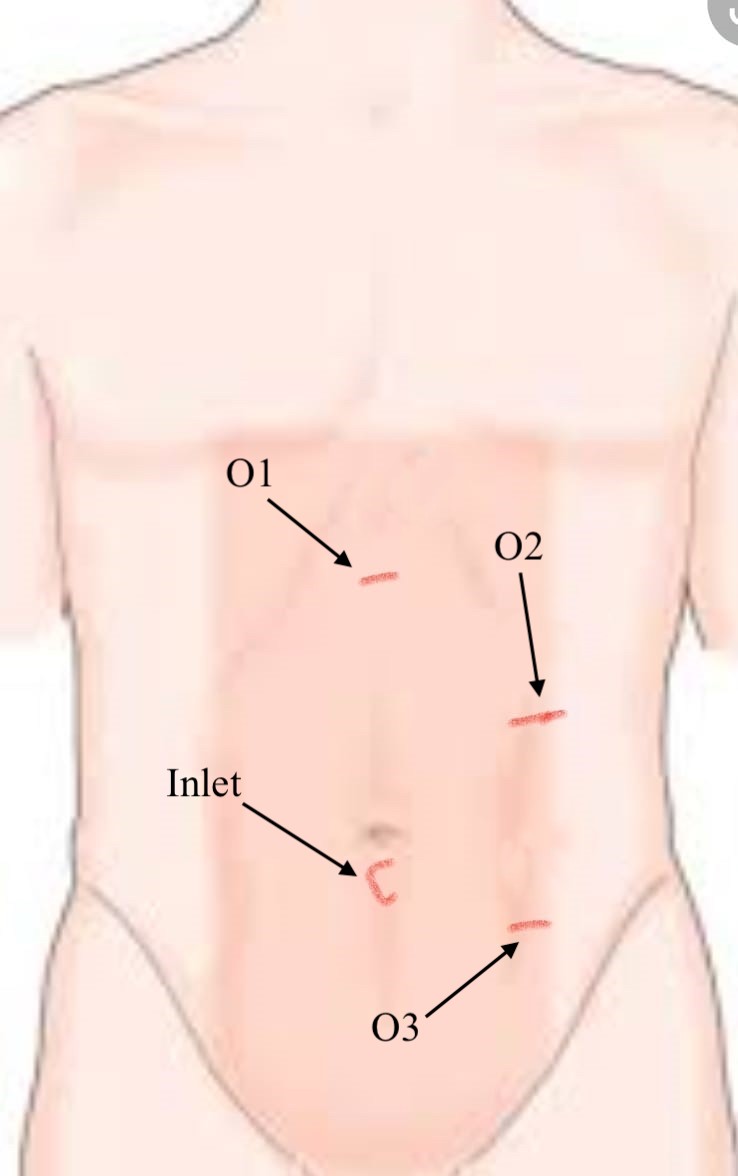}
		\caption{Case 3}
		\label{fig:Case3}
	\end{subfigure}
	\hfill	
	\begin{subfigure}[b]{0.2\textwidth}
		\centering
		\includegraphics[width=\textwidth]{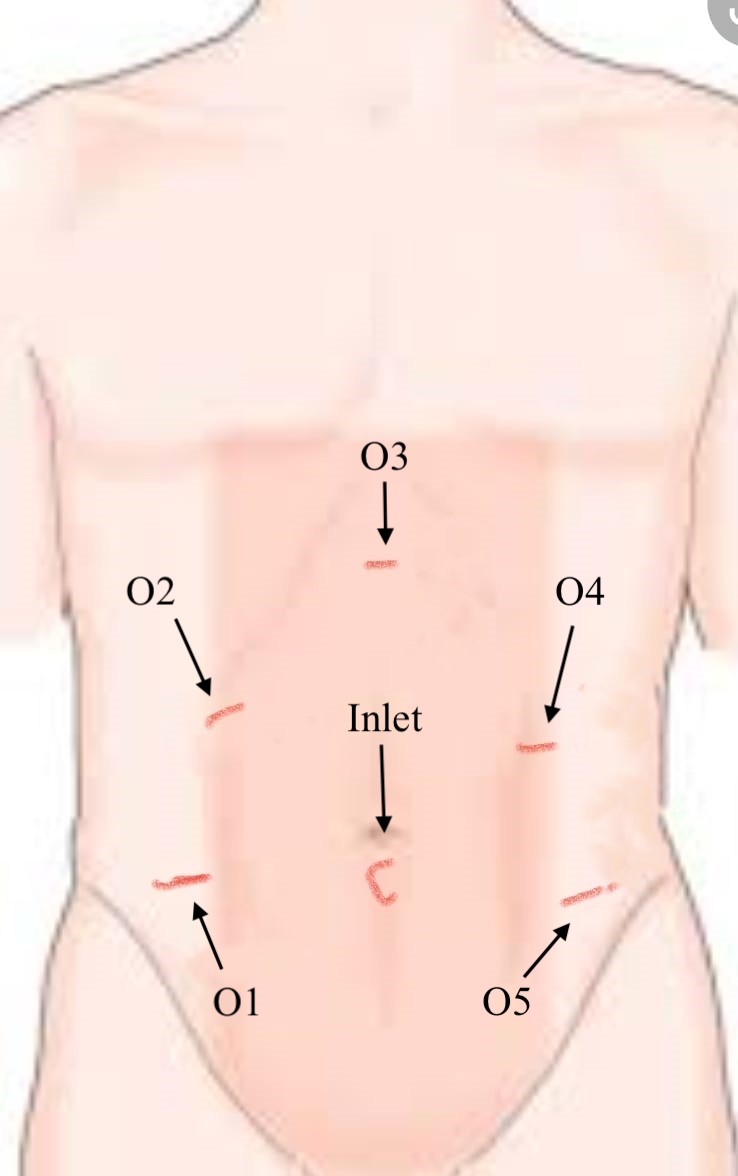}
		\caption{Case 4}
		\label{fig:Case4}
	\end{subfigure}
\caption{Four different types of laparoscopic surgery port configuration with different configurations of inlet and outlets trocars. The camera port, marked with `c', is typically used to supply $CO_2$ is placed at the navel.}
\label{fig:Exp-set-up}				
\end{figure}

Depending on the condition of the patient and the type of operation, the laparoscopic surgery set-up varies; nonetheless, the most common settings are the ones shown in Figure \ref{fig:Exp-set-up} which we will also consider in this work. Here, Case 1 represents laparoscopic cholecystectomy, where through the camera port, represented by `c', insufflation takes place and the rest of the three ports are used for the surgical tools, for instance, in line with the `c' port is the epigastric port through which the diathermy is used for the dissection purpose. Similarly, Cases 2 and 3, correspond to the setups for laparoscopic appendicectomy and laparoscopic hemicolectomy (right-sided colon surgery), respectively, where the energy device would be used from any of the two and three ports, respectively, depending on the operative stage and subsequent steps and finally, Case 4 is the left-sided colon resection or rectal surgery (see \cite{johnson1997laparoscopic} for more information). The diameter size of the ports varies from 5 mm to 12 mm, where a 12 mm port (typically only one) is used for the stapling device which is larger than the other instruments. A typical setup for Case 1 is shown in the left-hand side of Figure \ref{fig:MeshAndExpSetup}, and with this, our first goal is to develop a model that describes the dynamics of the flow inside the abdomen. 

For modelling purpose, the abdomen surface in Figure \ref{fig:Exp-set-up}, in its inflated form in a simplified way can be compared with a semi-ellipsoid. The corresponding mesh for this shape has been generated using \texttt{Gmsh v4.4.1} and the measurements correspond to those of an average human body. Next, an appropriate system of mathematical equations is defined and solved numerically. One commonly used turbulence model in CFD simulations is the $k-\omega$ SST (Shear Stress Transport) model \cite{menter2003ten}. It combines the advantages of the $k-\epsilon$ and $k-\omega$ turbulence models, making it suitable for a wide range of flow conditions. In particular, it is robust near boundaries and uses $k-\omega$ in the inner region, thus, well-suited for boundary layer flows, such as those encountered in laparoscopic surgery, where accurate prediction of the near-wall behaviour is crucial. The $k-\omega$ SST model provides enhanced accuracy and improved predictions of flow separation, making it a popular choice for simulating complex flow phenomena at a relatively low computational cost.

The computations and analyses are performed using OpenFOAM \texttt{v9}, an open-source software library written in C++ comprising many CFD solvers and utilities \cite{weller1998tensorial}. Several OpenFOAM utilities have been used for the simulations (e.g., mesh generation, parallelization, etc.) and the initial postprocessing (e.g., data visualization, calculation of field numbers, etc.). The standard solvers in OpenFOAM employ finite volume discretization to solve the governing equations. In our case, due to the turbulent nature of the underlying phenomenon, we use the PimpleFoam solver which discretizes and solves the momentum, continuity and energy equations (Navier--Stokes equations) in each cell. PimpleFoam is a large time-step transient solver for incompressible flow and uses the PIMPLE (merged PISO-SIMPLE) algorithm. 

\subsection{Flow model}
We consider a 3D, Newtonian and incompressible flow and implement a computational model based on the incompressible Navier--Stokes equations with turbulence. The flow domain representing an insufflated abdomen in a simplified way is taken as a semi-ellipsoid and it has a flat base. Thus, a system of partial differential equations, i.e., the continuity equation and the momentum equation for the conservation of mass and momentum, respectively and energy equations including a specific turbulence model is considered ({see Appendix for the equations and parameter values}).

\subsection{Mesh generation and CFD simulations}
The mesh has been generated using \texttt{Gmsh v4.4.1} and the geometry corresponds to the upper half of the spheroid with a minor axis of length 0.32m and a major axis of length 0.48m. The inlet and outlets of radii  5mm are placed as shown in Figure \ref{fig:Exp-set-up} and to have a finer mesh around them, \texttt{transfinite Curve}\footnote{The `Transfinite Curve' meshing constraints explicitly specifies the location of the nodes on the curve.} option with a value of 20 is used \cite{geuzaine2009gmsh}. The unstructured polyhedral mesh consists of predominantly tetrahedra finite volume cells and the quality of the mesh thus generated is assessed for the original configuration with the OpenFOAM utility \texttt{checkMesh}. With this, we compute the time evolution by solving the initial boundary value problem with the PimpleFoam solver with the $k-\omega$ SST model. The boundary conditions are given in Table \ref{tab:BC3D}, flow rates are of volumetric type with a value of inlet flow rate $V$ as 4.71 L/min. 

\begin{figure}
	\centering
	\begin{subfigure}[b]{0.3\textwidth}
		\centering
		\includegraphics[width=\textwidth]{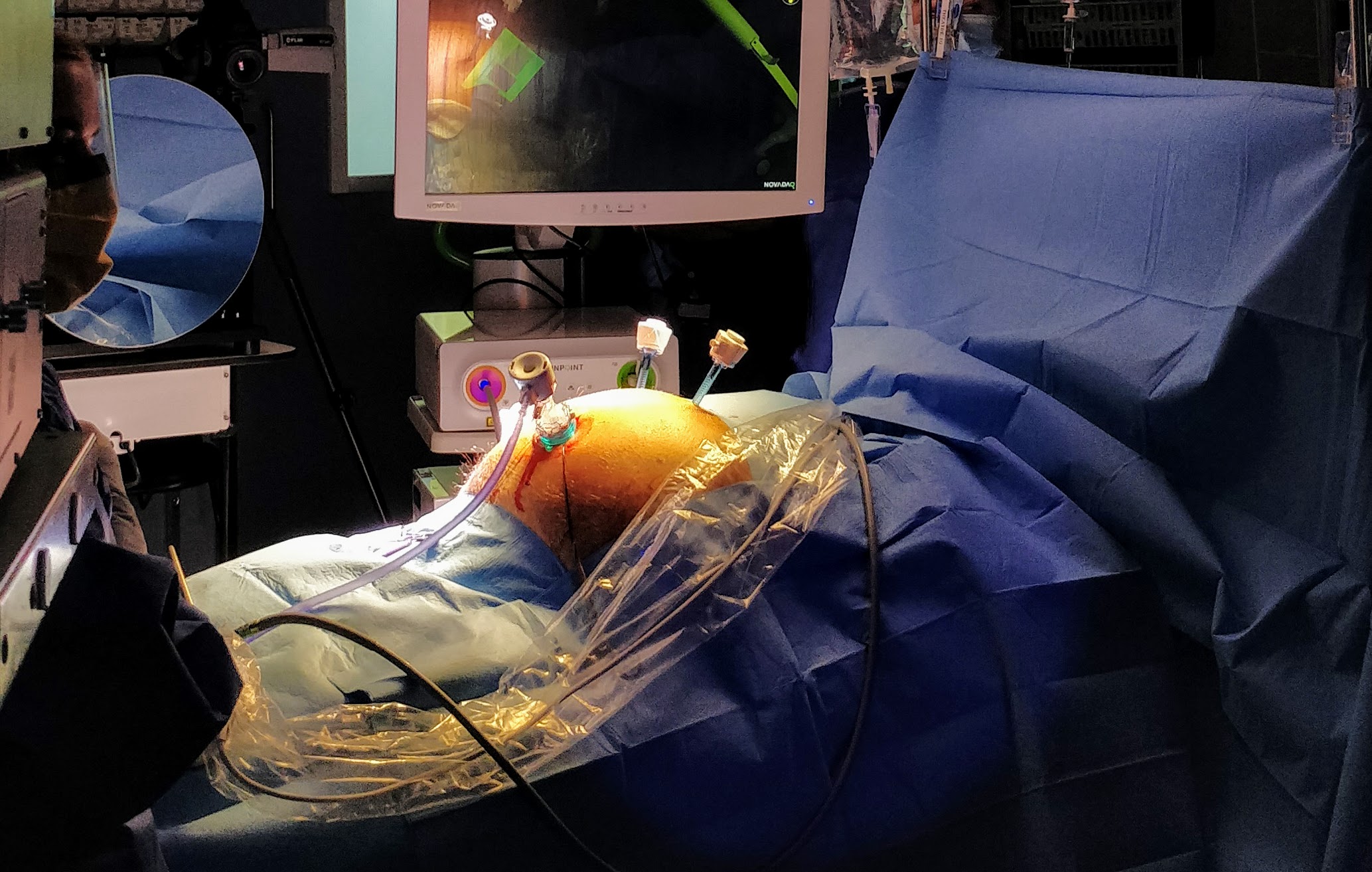}
		\caption{A laparoscopic cholecystectomy at Mater Misericordiae University Hospital, Dublin, Ireland.}
		\label{fig:ExpSetup}
	\end{subfigure}
	\hfill
	\begin{subfigure}[b]{0.3\textwidth}
		\centering
		\includegraphics[width=\textwidth]{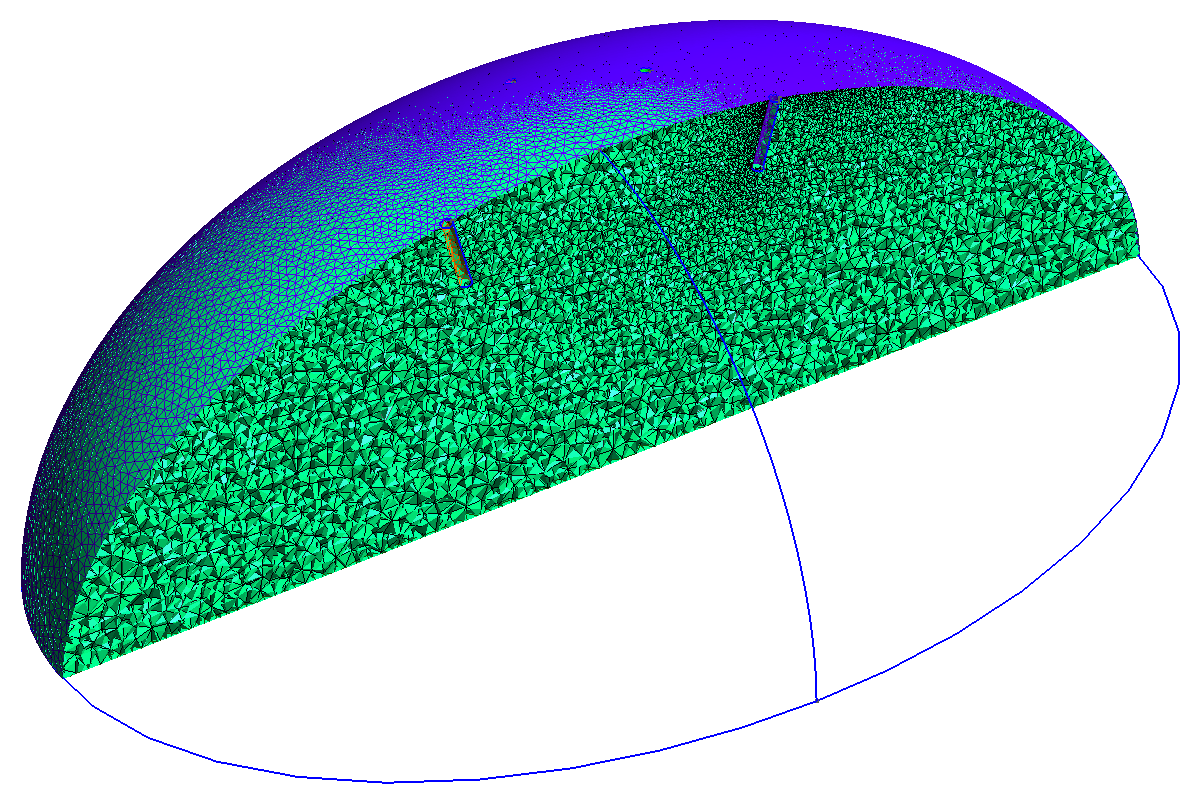}
		\caption{Mesh created in \texttt{Gmsh} for a simplified surgical domain.}
		\label{fig:Mesh3D}
	\end{subfigure}
	\begin{subfigure}[b]{0.3\textwidth}
	\centering
	\includegraphics[width=\textwidth]{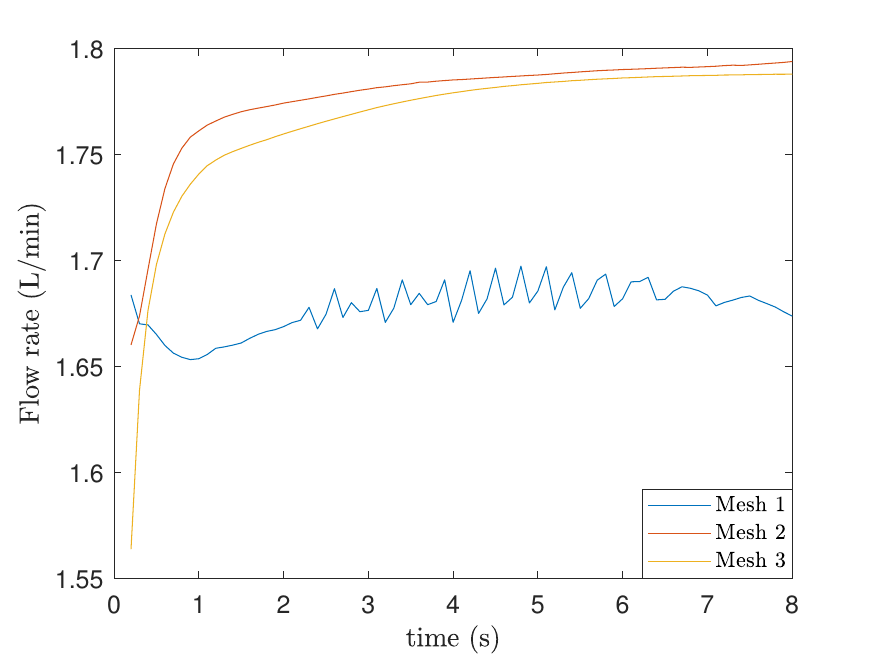}
	\caption{Flow rates with different meshes.}
	\label{fig:MeshIndep}
	\end{subfigure}
\caption{}
\label{fig:MeshAndExpSetup}
\end{figure}

To verify grid independence, we perform simulations on three meshes each of which has greater grid density. More precisely, for Case 1, the three meshes, mesh 1, mesh 2 and mesh 3, consist of 83208, 191888, and 550702 cells, respectively. To compare them, we calculate the flow rate for outlet 1 over the same simulation time and plot in Subfigure \ref{fig:MeshIndep} and notice a good agreement in the basic trends of the three curves. It can also be seen that the flow rate of mesh 1 and mesh 2 differ more than that of mesh 2 and mesh 3, and a similar pattern is observed for the other two outlets as well; thus, demonstrating the grid independence. For the rest of our computations, we have considered mesh parameters set for mesh 3. 

\begin{table}
	\centering

	\begin{tabular}{ |p{0.5cm}||p{4cm}|p{3.5cm} |p{4cm} |p{3cm} | }
		\hline 
		& inlet  & outlet & wall & internalField\\
		\hline		
		$\omega$ & fixedValue, uniform 440.15 & zeroGradient &  omegaWallFunction, uniform 440.15 & uniform 440.15  \\
		$\epsilon$ &  zeroGradient & zeroGradient & epsilonWallFunction, uniform 200 & uniform 200   \\		
		$u_j$ &  uniform, volumetricFlowRate $7.85\cdot 10^{-5}$ & zeroGradient & fixedValue, uniform 0 &  \\		
		$k$ &  turbulentIntensity\-KineticEnergyInlet, uniform 0.375 & zeroGradient & kqRWallFunction, uniform 0 & uniform 0.375  \\
		$\nu_t$ & calculated, uniform 0 & calculated, uniform 0 &  nutkWallFunction, uniform 0 &  uniform 0  \\
		$p$ &  zeroGradient & fixedValue, uniform 0 & zeroGradient & uniform 0  \\	
		\hline
	\end{tabular}
	\caption{The boundary conditions for the CFD model.}
	\label{tab:BC3D}	
\end{table}

\begin{figure}
	\centering
	\begin{subfigure}[b]{0.48\textwidth}
		\centering
		\includegraphics[width=\textwidth]{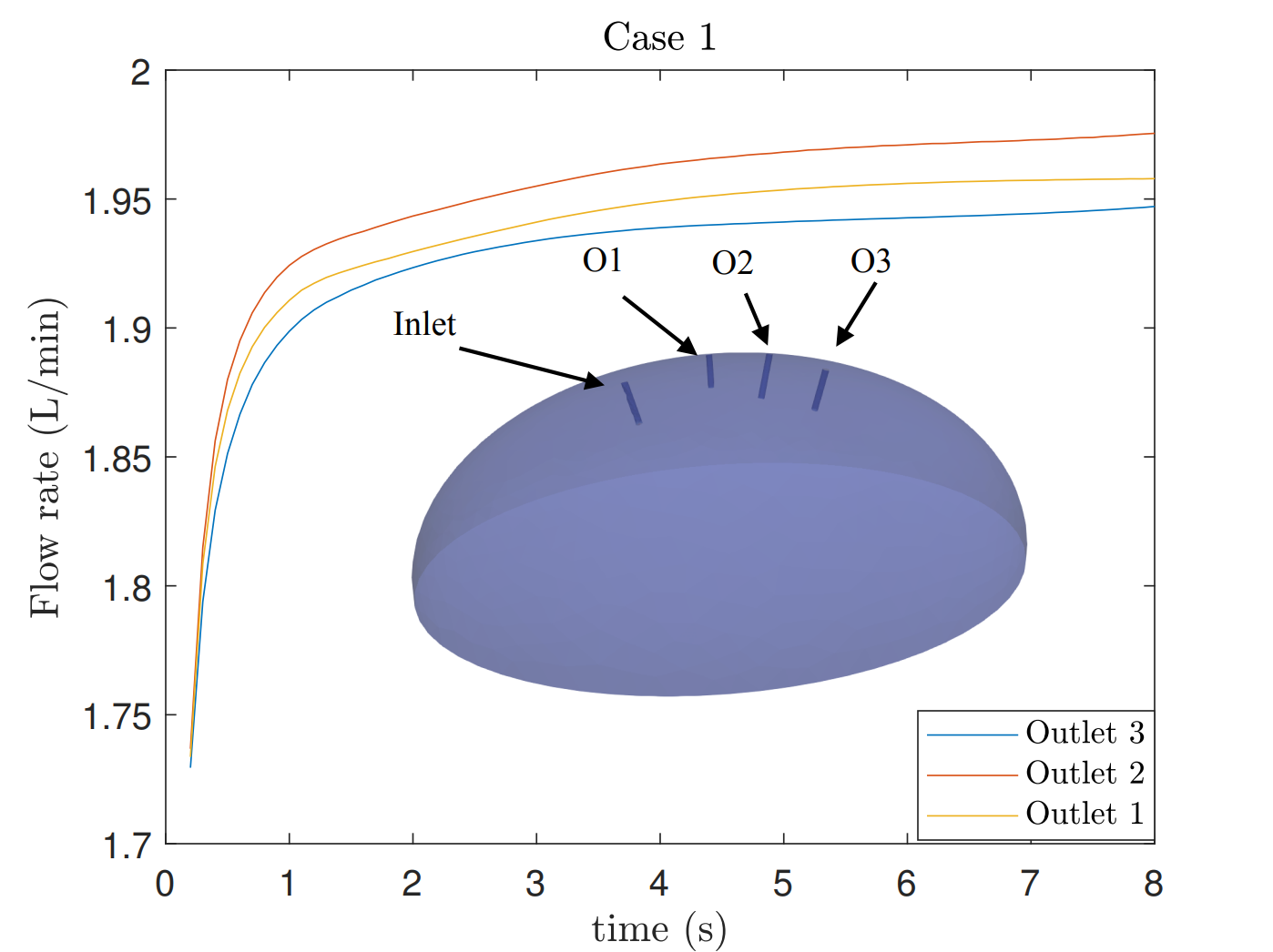}
		\label{fig:FlowRate1}
	\end{subfigure}
	\hfill
	\begin{subfigure}[b]{0.488\textwidth}
		\centering
		\includegraphics[width=\textwidth]{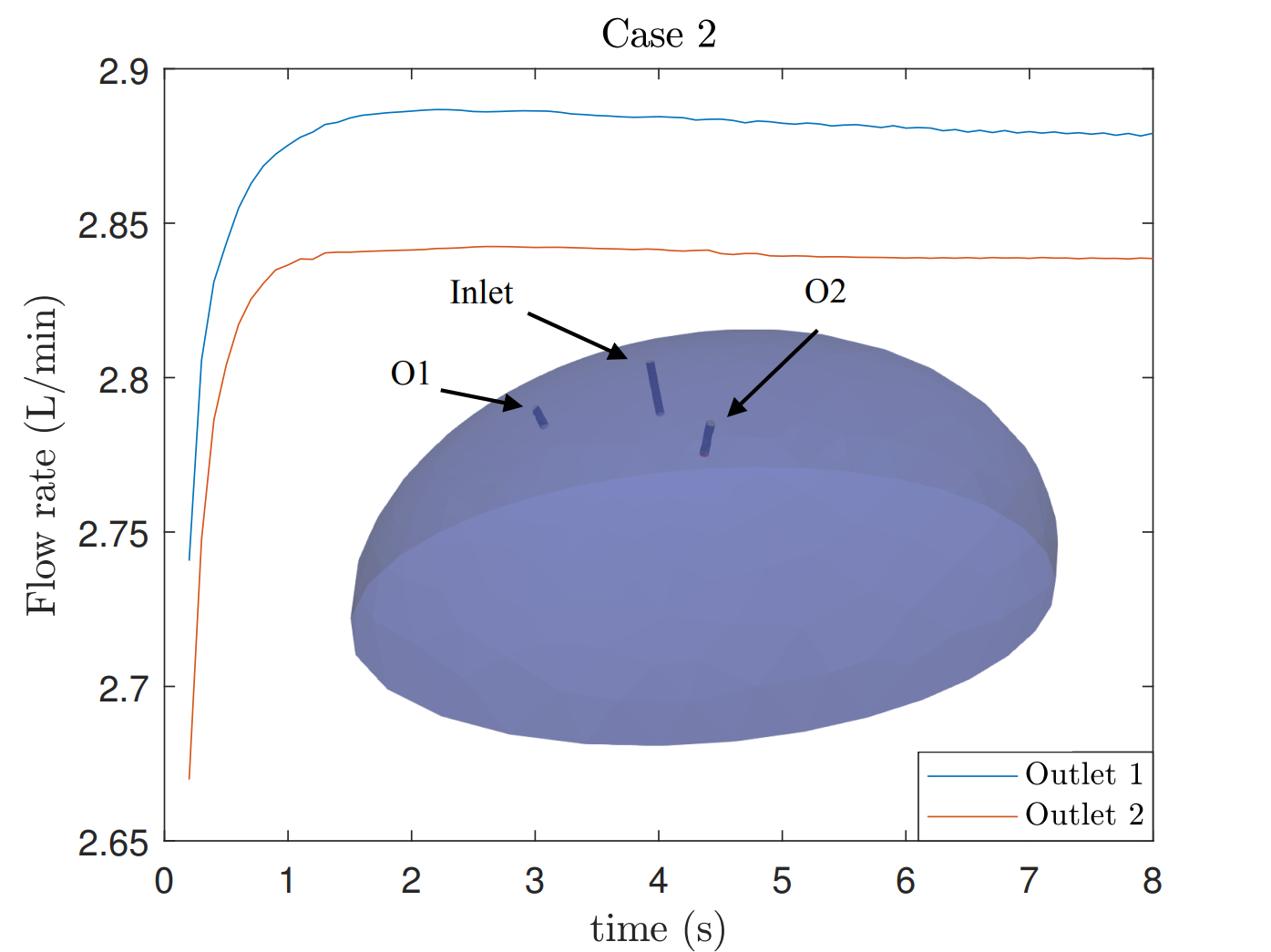}
		\label{fig:FlowRate2}
	\end{subfigure}
	\hfill
	\begin{subfigure}[b]{0.488\textwidth}
		\centering
		\includegraphics[width=\textwidth]{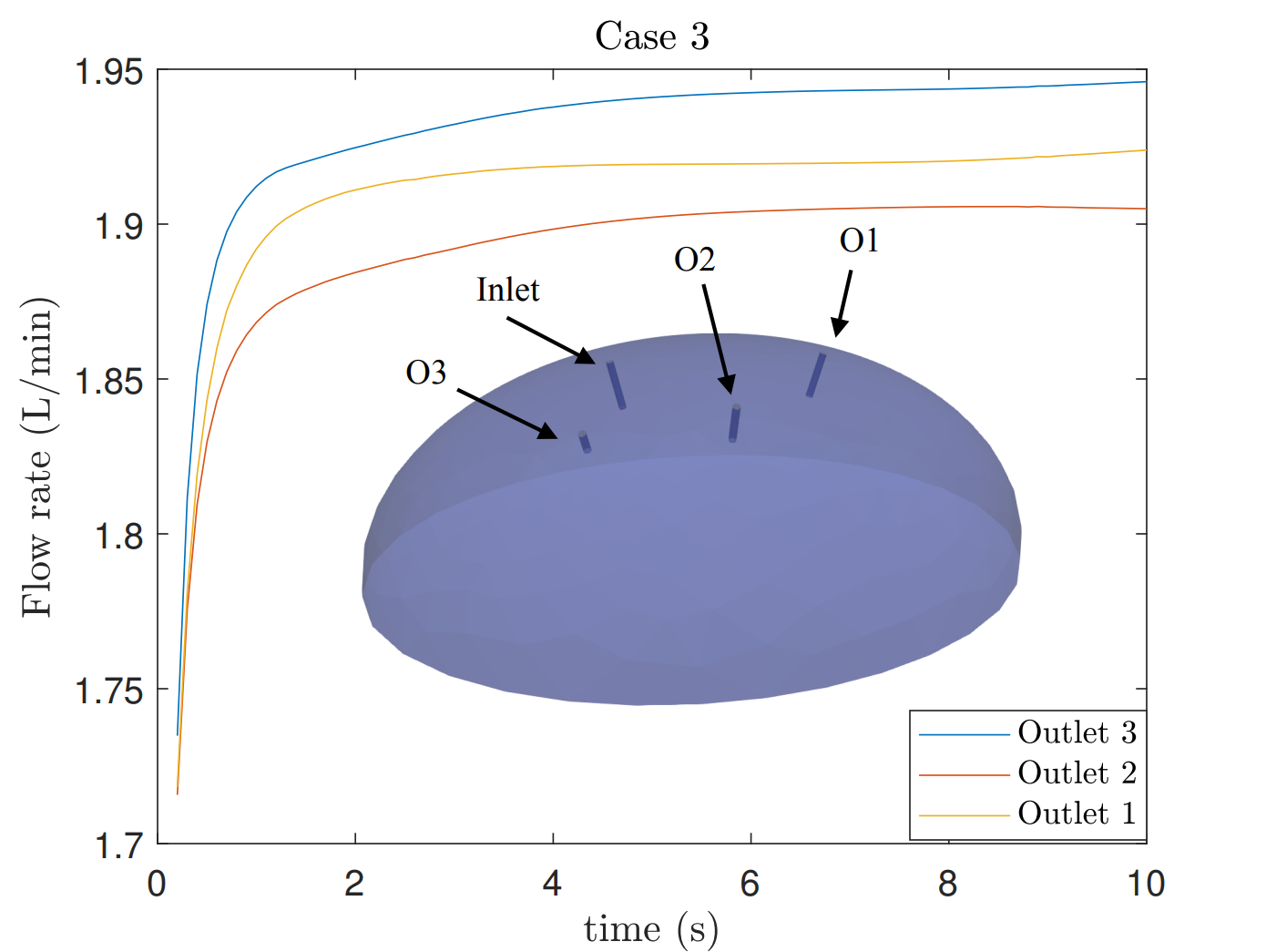}
		\label{fig:FlowRate3}
	\end{subfigure}
	\begin{subfigure}[b]{0.49\textwidth}
		\centering
		\includegraphics[width=\textwidth]{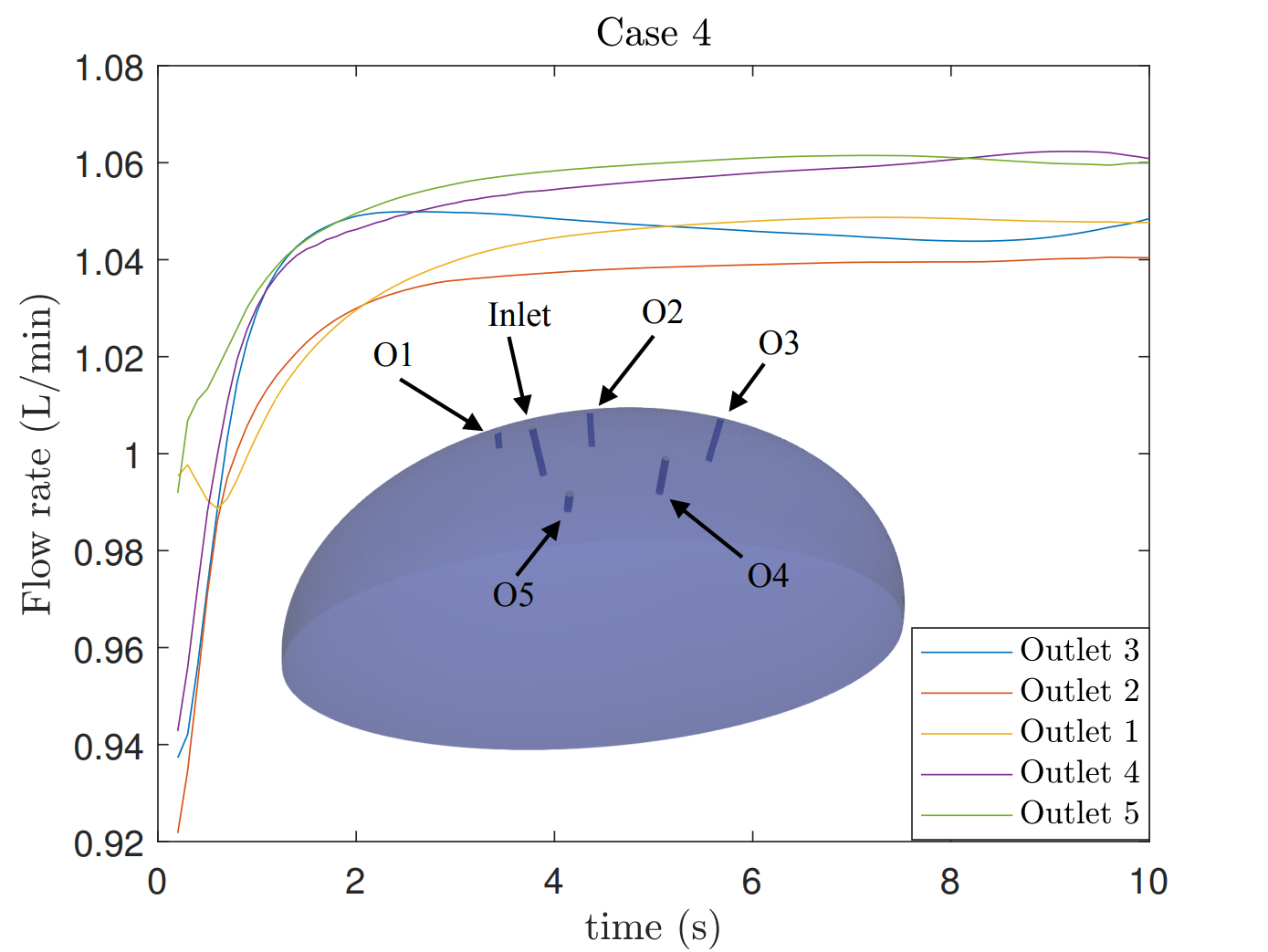}
		\label{fig:FlowRate4}
	\end{subfigure}
	\caption{Flow rates for each of the four cases shown in Figure \ref{fig:Exp-set-up} plotted on a semilogarithmic scale. Also shown is the corresponding domain geometry with inlet and different outlets indicated as O1, O2, representing Outlet 1, Outlet 2, etc., respectively.  Clearly, the flow rate depends on the position and number of the outlet trocars.}  
	\label{fig:FlowRates}	
\end{figure}

\section{Results and Discussion}
\label{sec:ResultsDiscussion}
With the parameters and methodology discussed above, we have calculated the numerical evolution for 10 seconds. We compute the flow rates as a function of time for each of the outlets for different cases as shown in Figure \ref{fig:FlowRates}. The different colours correspond to different outlets and their values indicate that the flow rates depend on the position and number of outlets. The numerical simulations (\href{https://youtu.be/xMFBq5DtMQc}{link}) show the evolution of smoke particles that follow the eddies generated after the injection of the gas. Indeed, the movement (transportation, mixing) of these particles is influenced by the underlying flow dynamics and to understand it, one can not always rely upon the instantaneous velocity field as the instantaneous streamlines can diverge from actual particle trajectories very quickly \cite{shadden2005definition}. Furthermore, these streamlines vary when viewed from different reference frames, for instance, a region with closed streamlines in one frame can appear completely different when viewed in another frame, and they do not serve for understanding the material transport \cite{haller2015lagrangian}. Thus, instead of an Eulerian perspective, we work in a Lagrangian framework and for a given time interval, we calculate the particle trajectories and quantities that are also frame-invariant, and reveal the flow structures such as circulation, vortex formation as discussed in the following. 
\subsection{Lagrangian Coherent Structures and their application}
Given a two-dimensional velocity field $\mathbf{u}(\mathbf x, t)$ of the form
\begin{equation}
\label{eq:xprime}
\mathbf{x}^\prime = \mathbf {u} (\mathbf x, t), \ \mathbf{x}\in \Omega \subset \mathbb{R}^2, \ t\in[t_-, t_+],
\end{equation}
the trajectories it generates are denoted by $\mathbf{x}(t; \mathbf{x}_0, t_0 )$, with a position $\mathbf x_0$ at time $t_0$. The dynamics of fluid elements can be explained with the flow map
\begin{equation}
F^t_{t_0}(\mathbf x_0) \equiv \mathbf x(t; t_0, \mathbf x_0),
\end{equation}
mapping initial positions $\mathbf x_0$ to current positions at time $t$. Using the flow map, the right Cauchy--Green strain tensor field $C_{t_0}^t(\mathbf x_0)=\nabla F_{t_0}^t(\mathbf x_0)^T \nabla F_{t_0}^t(\mathbf x_0)$ is defined which characterizes Lagrangian strain in the flow. It is symmetric and positive definite and thus, its eigenvalues $\lambda_j(\mathbf x_0)$ and eigenvectors $\xi_j(\mathbf x_0)$ satisfy 
\begin{align}
\label{eq:CauchyGreen}
	C_{t_0}^t \xi_j &= \lambda_j \xi_j, \, j = 1, 2; 0<\lambda_1\leq\lambda_2,	\\
	|\xi_j|&=1, \, \xi_2 = \Omega \, \xi_1, \, \Omega=
	\begin{pmatrix}
	0 & -1\\ 1 & 0
	\end{pmatrix}.		
\end{align} 
A particle initially located at $\mathbf x_0$ at $t_0$, when advected, moves to $F_{t_0}^{t_0+\tau}$ after a time interval $\tau$ and the amount of stretching about this trajectory can be characterized by the finite time Lyapunov exponent (FTLE) defined as
\begin{equation}
\label{eq:FTLE-def}
FTLE(\mathbf x_0, t_0) = \frac{1}{|\tau|}\ln \sqrt{\lambda_{\max}(C_{t_0}^{t_0+\tau})},
\end{equation}
where $\lambda_{\max}$ is the largest eigenvalue of $C_{t_0}^{t_0+\tau}$. In other words, it describes a finite time average of the maximum expansion rate for a pair of particles advected by the flow \cite{shadden2005definition}. 

The FTLE field is both space and time-dependent and ridges of local maxima in the field represent material lines in the flow and reveal transport barriers between the different regions of the flow. 
In \eqref{eq:FTLE-def}, an absolute value of $\tau$ is used since the FTLE can be computed for forward times ($\tau>0$) and backward times ($\tau<0$). If the FTLE field is calculated by integrating trajectories in backward time, the fluid particles tend to collect or accumulate in local structures and the corresponding ridges are called attracting (unstable) material lines. On the other hand, for the forward time, ridges correspond to the repelling (stable) lines of maximum spreading so that particles initially close, diverge quickly in forward time. These separatrices are called Lagrangian Coherent Structures (LCS) and are analogues of stable and unstable manifolds in the time-dependent systems \cite{haller2015lagrangian}. Moreover, they divide dynamically distinct regions in the flow and reveal geometry which is often hidden when viewing the vector field, or even trajectories of the system \cite{haller2001distinguished}. 
\begin{figure}[h]
	\includegraphics[width=0.4\textwidth]{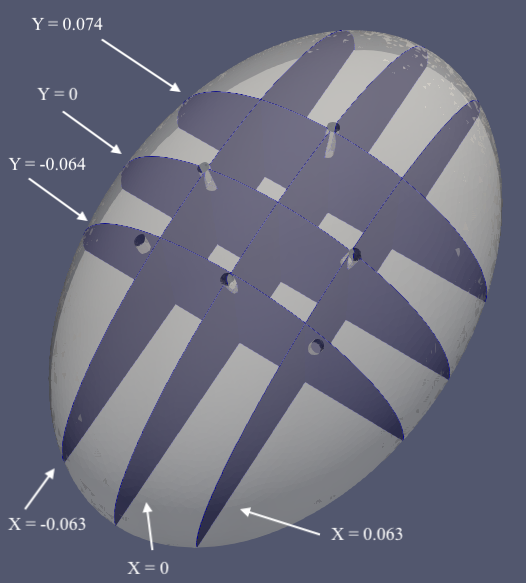}
	\caption{Different cross-section planes of Case 4.}
	\label{fig:Case4Allplanes}
\end{figure}

To calculate FTLE numerically, the deformation gradient tensor needs to be estimated and to obtain that, fluid particle path positions are extracted over a finite time interval from CFD simulations described in the last section. Following the procedure \cite{shadden2005definition}, particle trajectories $\mathbf{x}(t)$ are determined numerically for the time period $\tau$, by solving \eqref{eq:xprime} using the Runge--Kutta (4,5) method. To obtain the velocity along particle trajectories, a linear interpolation scheme is employed in space and time and a free slip boundary condition is used and the calculation of the trajectory of particles is stopped once they exit the flow domain.  
 
Among the four cases of laparoscopic surgery, the flow dynamics in Case 4 is expected to be more complicated due to the number of outlets, although the computations have been performed for the other three cases as well, the observations made are similar, and therefore, for brevity, we present our results for this case. Moreover, as our goal is to visualize the FTLE field and detect the LCSs by plotting them, we achieve this by considering different planar cross-section areas. We perform the FTLE computations for a two-dimensional velocity field for six different planes located at $X=-0.063, 0, 0.063$ and $Y=-0.064, 0, 0.074$ as shown in Figure \ref{fig:Case4Allplanes}. Each of these planes intersects with at least two trocar ports, except the one at $Y=0.074$, that contains only one port in three out of the four surgical cases, while in Case 2 there is no trocar present in that section. For the calculation of the FTLE field we first calculate the velocity field over a uniform grid with step size $\Delta x = \Delta y = 3\cdot 10^{-3}$, using a linear interpolation and then the computations for the FTLE field are performed on a finer grid with step size $\Delta x=\Delta y=10^{-3}$.
\begin{figure}
\begin{subfigure}[b]{.48\linewidth}
	\centering
	\includegraphics[width=\linewidth]{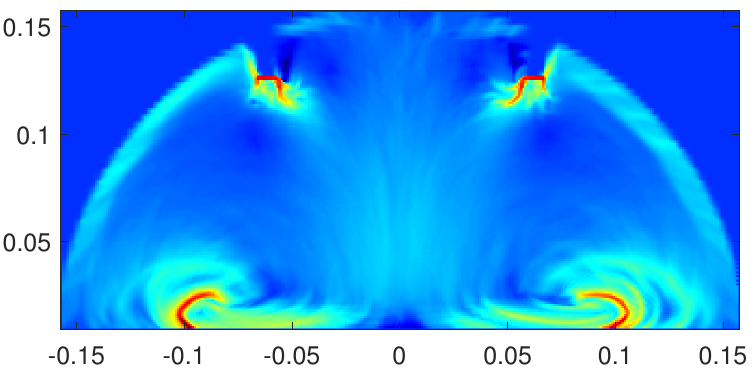}	
	\caption{$t\in[1,3]$}
	\label{fig:FTLEt2}
\end{subfigure}\hfill
\begin{subfigure}[b]{.48\linewidth}
	\centering
	\includegraphics[width=\linewidth]{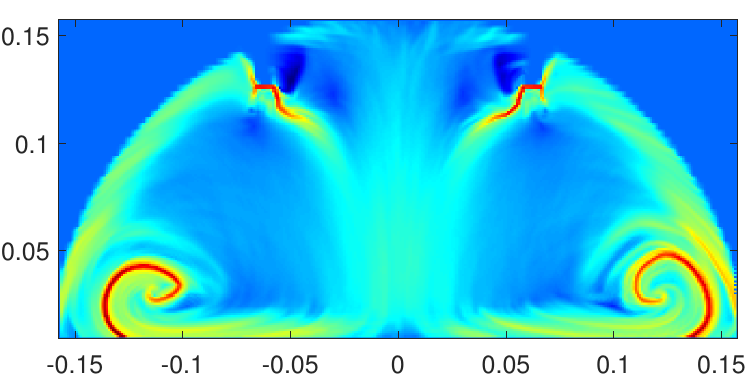}
	\caption{$t\in[1,5]$}		
	\label{fig:FTLEt4}
\end{subfigure}
\begin{subfigure}[b]{.48\linewidth}
	\centering
	\includegraphics[width=\linewidth]{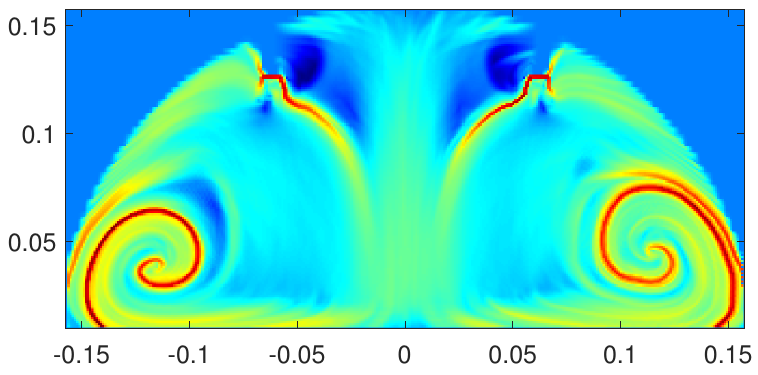}	
	\caption{$t\in[1,7]$}		
	\label{fig:FTLEt6}
\end{subfigure}\hfill
\begin{subfigure}[b]{.48\linewidth}
	\centering
	\includegraphics[width=\linewidth]{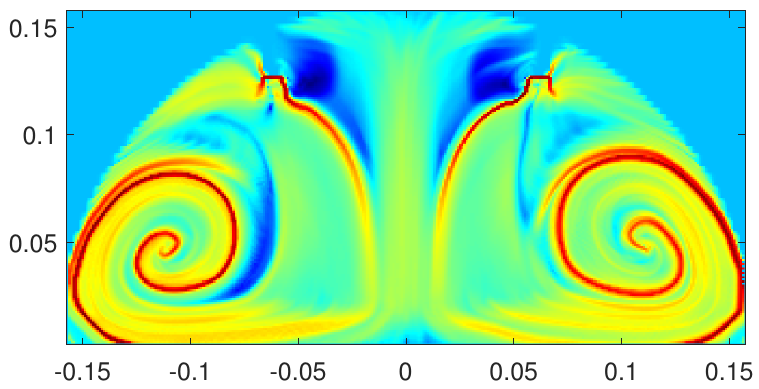}	
	\caption{$t\in[1,9]$}		
	\label{fig:FTLEt8}
\end{subfigure}
\caption{Dependence of FTLE field on the time interval: Backward FTLE field calculated for Plane $Y=0$ of Case 4 for different time periods.}
\label{fig:FTLEdifftimeperiod}
\end{figure}	
\subsubsection{The backward FTLE field and its dependence on the time interval}
For calculating the FTLE field, one of the important factors is the integration time period $\tau$ over which the trajectories are calculated. An adequate choice is also crucial to observe the fluid flow structures which are responsible for the dispersion or accumulation. In this work, one of our main goals is to understand the accumulation of smoke particles, and for that, we examine the backward FTLE field. Thus, for the plane $Y=0$, we have calculated it for four different time periods, i.e., $\tau=2, 4, 6, 8$ seconds, as shown in Figure \ref{fig:FTLEdifftimeperiod}, where the FTLE field values are normalized. The red part highlights the areas with the maximum value of the FTLE field and the blue with the minimum value, and it can be noticed that with a longer time period, more details in the field are revealed. These high values correspond to the ridges of the FTLE field which, in turn, form the LCSs, the attracting material lines. When the time period is 2 seconds these lines are small in size, which progressively increases with the time period. As a longer time period also imply higher computational cost, we choose a moderately large value, i.e., 6 seconds for the rest of the computations. 

\begin{figure}[h]
	\includegraphics[width=0.7\linewidth, valign=c]{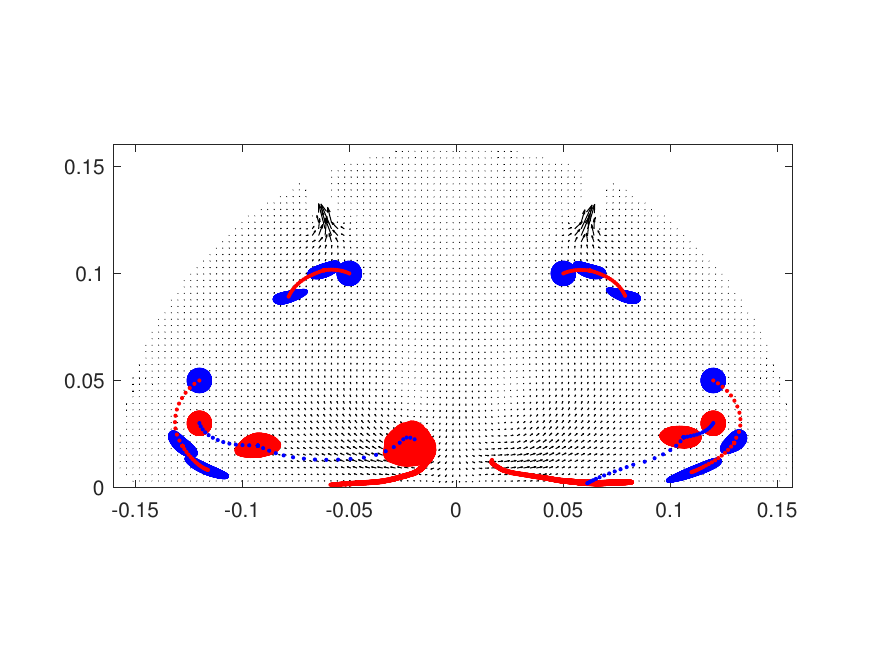}\\
	\caption{The velocity field in the plane $Y=0$ and the backward movement of tracer particles from the time $t=5$, when they are in the form of a circle, to time $t=3$, i.e., the half time period, to time $t=1$, the final time. The trajectory of one of the particles is also highlighted for all the times steps in between.}
	\label{fig:backFTLEandTracers}
\end{figure}

In Figure \ref{fig:backFTLEandTracers}, we display the movement of tracer particles plotted over the corresponding velocity field, which also explains the role of LCSs. We consider $10^{4}$ tracer particles in the form of a circle and at $t=5$, we place them on each side of the LCS curve that appears near the bottom, shown in the red and blue colour, respectively. The idea is to observe their movement backwards in time, i.e., starting at time $t=5$, until $t=1$ second; the same is done for the LCS curve on appearing on the opposite side. The evolution is shown at time $t=5$, $t=3$, and $t=1$, and the trajectory of a single particle is displayed for all the times in between and is marked with a colour opposite to the rest to visualize it. The plot shows that the two particles away from each other at $t=1$, tend to come closer at $t=5$, i.e., near the LCS curves, thus, confirming the presence of attracting lines in that region. This behaviour can be observed for the LCS lines in the top region close to the outlets where we have placed another set of tracer particles. This observation in the context of surgical smoke is useful as it informs the location where the possible accumulation could take place. 

\subsubsection{Effect of the spatial position on the FTLE field}		

\begin{figure}
	\begin{subfigure}[b]{.52\linewidth}
		\centering
		\includegraphics[width=\linewidth]{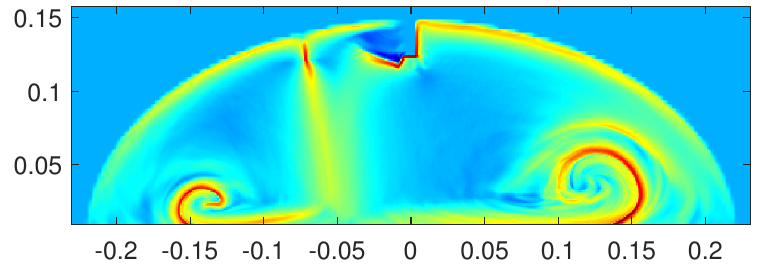}		
		\caption{Case 4, $X=-0.063$}
		\label{fig:Case4Xm063}
	\end{subfigure}\hfill
	\begin{subfigure}[b]{.42\linewidth}
		\centering
		\includegraphics[width=\linewidth]{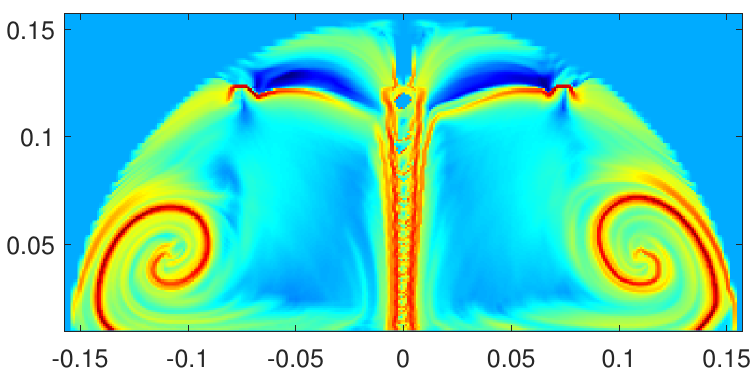}		
		\caption{Case 4, $Y=-0.064$}		
		\label{fig:b}
	\end{subfigure}\\
	\begin{subfigure}[b]{.52\linewidth}
		\centering
		\includegraphics[width=\linewidth]{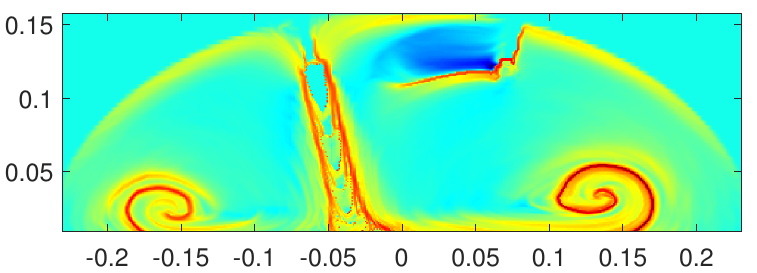}	
		\caption{Case 4, $X=0$}
		\label{fig:a}
	\end{subfigure}\hfill
	\begin{subfigure}[b]{.42\linewidth}
		\centering
		\includegraphics[width=\linewidth]{bFTLECase4Y0len60t7}	
		\caption{Case 4, $Y=0$}
		\label{fig:b}
	\end{subfigure}\\
	\begin{subfigure}[b]{.52\linewidth}
		\centering		
		\includegraphics[width=\linewidth]{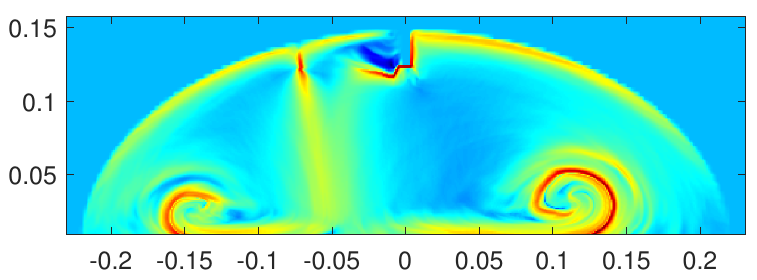}
		\caption{Case 4, $X=0.063$}
		\label{fig:Case4X063}
	\end{subfigure}\hfill
	\begin{subfigure}[b]{.42\linewidth}
		\centering
		\includegraphics[width=\linewidth]{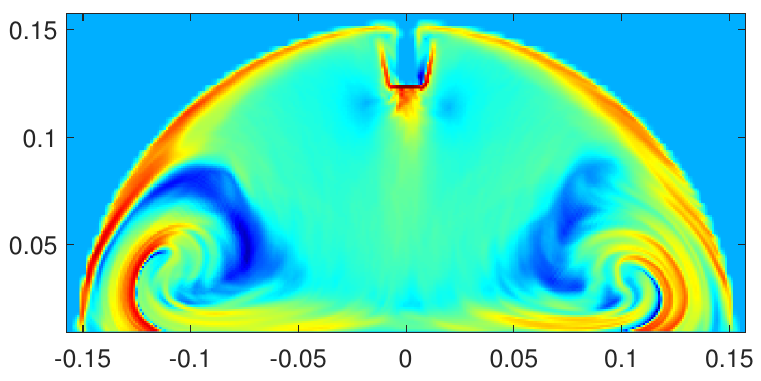}		
		\caption{Case 4, $Y=0.074$}
		\label{fig:b}
	\end{subfigure}\\
	\begin{subfigure}[b]{.52\linewidth}
		\centering		
		\includegraphics[width=\linewidth]{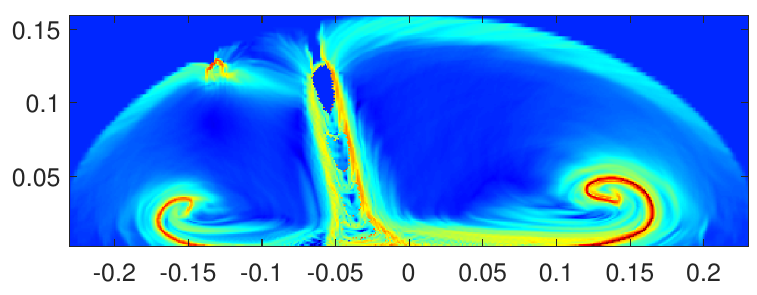}
		\caption{Case 2, $X=0$}
		\label{fig:Case2X0}
	\end{subfigure}\hfill
	\begin{subfigure}[b]{.42\linewidth}
		\centering
		\includegraphics[width=\linewidth]{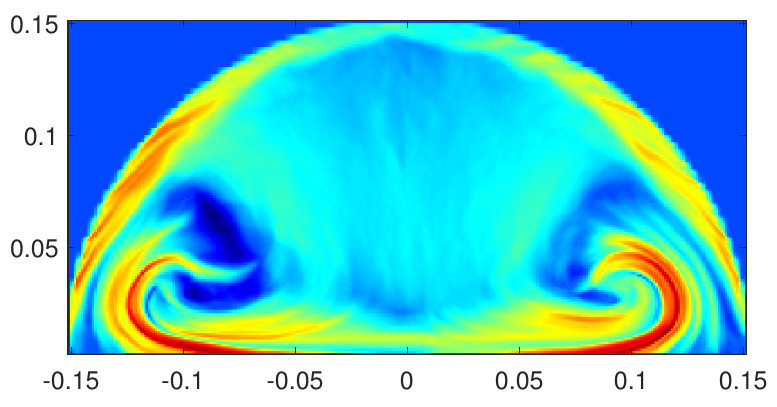}	
		\caption{Case 2, $Y=0.074$}
		\label{fig:Case2Y074}
	\end{subfigure}\\

	\caption{The backward FTLE field for time period $t\in[1,7]$ in different cross-sectional planes of Case 4, and Case 2.}
	\label{fig:1}
\end{figure}

The flow rate calculations in the earlier section clearly show that the flow dynamics would change with the location and number of outlets. We analyse this fact further by considering different plane configurations as mentioned above. In Subfigure \ref{fig:Case4Xm063}, we consider the plane $X=-0.063$, containing outlets 1 and 2, and following the ridges, we observe vortex formations mainly in two areas. Furthermore, the areas near the outlets also have a large FTLE value implying a possible site of accumulation. A similar behaviour is observed when we look at the plane $X=0.063$, containing outlets 4 and 5, as in Subfigure \ref{fig:Case4X063}, for example, the location and size of ridges are comparable. However, if we consider the plane $X=0$, which contains an inlet and outlet 3, the ridges unveiling two vortices are more prominent. Also, the ridges are nearly straight lines inclined at an angle made by the inlet; they clearly divide the flow into two different parts which implies that the particles on one side remain that side and do not cross to the other side, thus not allowing any mixing. This cross-sectional plane is crucial as outlet 3 is the epigastric port through which the instruments are placed for dissection. This jet impingement behaviour caused by the inlet can be observed also if we look at the plane $Y=-0.064$, containing the inlet and two outlets, where the vertical ridges are formed dividing the flow domain into two different regions, each containing a vortex dictated by the ridges. These vortices continue to be present as we move away to the plane $Y=0$, which contains two outlets; nonetheless, the ridges near the outlets do deform from their previous shape. As we move further away to the plane $Y=0.074$, where only one outlet is present, the FTLE field is quite different from the previous two cases. With one outlet, not only the vortex formations are less prominent, but the abdomen wall area also has a large FTLE value implying a possibility of accumulation. This behaviour is observed in Cases 1 and 3, except that in Case 2, the accumulation near wall decreases as shown in Subfigure \ref{fig:Case2Y074}, which can be justified with the outlet present in the plane in the other three cases. Finally, for the sake of completeness, we also present the FTLE field for the plane $X=0$ in Case 2, which is different from the other three cases as the outlet is on the left side to the inlet. The results are consistent with the previous observations, i.e., besides the almost straight ridges, there are vortex formations in the bottom region. 

\section{Conclusions}
\label{sec:Conclusions}

In this work, our primary objective has been to gain insight into the behaviour of surgical smoke, in particular, the accumulation and mixing, during laparoscopic surgery. To achieve this, we adopt a Lagrangian framework, which allows us to study the underlying flow more accurately compared to traditional Eulerian techniques.

To analyze the flow, we calculate the velocity field in a semielliptical-shaped geometry using a CFD model and the flow rate calculations serve as a basis for experimental validations of our numerical results. Additionally, we employ the Finite Time Lyapunov Exponent (FTLE) field to further investigate the flow dynamics. The backward  FTLE ridges reveal intricate and complex flow patterns, which enable us to identify regions of vortex formation and maximum accumulations in different planar regions of the abdomen.

Our investigation is particularly motivated by the need to remove smoke before it obstructs visualization during surgery and to limit patient exposure. By studying the FTLE field, we gain insights into inflow jet impingement and the resulting vortex behaviour, which govern the transport, mixing, and accumulation of the smoke material. This knowledge is crucial as it can guide the effective placement of instruments for the efficient removal of surgical smoke. While our LCS analysis focuses on a simplified scenario with a flat surface in the abdomen domain, the striking observations we have made hold significant potential for real-life surgical scenarios. Although an implementation in OR requires further analysis of experimental data, which we plan to address in future, the current findings provide a novel approach to the applicability of LCS in enhancing surgical procedures and contributing to a safer and more efficient surgical environment.

\section*{Acknowledgments}
This work has been supported by the project PORSAV (Protecting Operating Room Staff Against Viruses) funded by the European Union’s Horizon 2020 research and innovation programme under grant agreement No 101015941. The authors wish to acknowledge the Irish Centre for High-End Computing (ICHEC) and the ResearchIT Sonic cluster for the provision of computational facilities and support. 

\section*{Appendix}

By writing the fluid velocity and the position coordinates in tensor form as $u_j$ and $x_j$, respectively, with direction $j$, the continuity equation is given by

\begin{equation}
	\label{eq:conteq}
	\frac{\partial u_j}{\partial x_j} = 0.
\end{equation}
Similarly, the momentum equation can be written as
\begin{equation}
	\frac{\partial}{\partial t} (\rho u_i) + \frac{\partial}{\partial x_j} (\rho u_i u_j) = - \frac{\partial p}{\partial x_i} + \frac{\partial \tau_{ij}}{\partial x_j},
\end{equation}
where $\rho$ is the fluid density, $p$, static pressure and the shear stress $\tau_{ij}$, i.e., $j^{th}$ component of the stress acting on the faces of the fluid element perpendicular to axis $i$, is dependent on the fluid viscosity $\mu$ by
\begin{equation}
	\tau_{ij} = \mu \left(\frac{\partial u_i}{\partial x_j} +\frac{\partial u_j}{\partial x_i} \right).
\end{equation}
On the other hand, the energy equations are derived using the $k-\omega$ SST turbulence model which combines the best of the $k-\omega$ model and $k-\omega$ model with a high Reynolds number. Hence, the turbulent kinetic energy $k$ and dissipation rate $\omega$ are modelled by different equations where the transport equation for $k$ is
\begin{equation}
	\frac{\partial (\rho k)}{\partial t} + \frac{\partial (\rho u_ik)}{\partial x_i} = \frac{\partial }{\partial t} \left( (\mu + \sigma_k \mu_t) \frac{\partial k}{\partial x_i}\right) + \tilde P_k - \beta^\star \rho \omega k,
\end{equation}
where the left-hand side corresponds to the time derivative and convection terms for $k$ and the right-hand side has the diffusion, production and dissipation terms, respectively. In particular, a production limiter is employed to avoid the build-up of turbulence in the stagnation regions with
$$
P_k = \mu_t \frac{\partial u_i}{\partial x_j} \left(\frac{\partial u_i}{\partial x_j} + \frac{\partial u_j}{\partial x_i} \right), \, \tilde{P}_k = \min(P_k, 10\cdot \beta^* \rho k \omega).
$$
Similarly, the transport equation for $\omega$ is
\begin{equation}
	\frac{\partial (\rho \omega)}{ \partial t } + \frac{\partial (\rho u_i \omega)}{\partial x_i} = \frac{\partial }{\partial x_i} \left( (\mu + \mu_t \sigma_\omega ) \frac{\partial \omega}{\partial x_i}\right) + \alpha \rho S^2 - \beta \rho \omega^2 + 2(1-F_1) \frac{\rho \sigma_{\omega_2}}{\omega} \frac{\partial k}{\partial x_j} \frac{\partial \omega}{\partial x_j}, 
\end{equation}
where the blending function 
$$
F_1 = \tanh\left(\left(\min \left(\max \left(\frac{\sqrt{k}}{\beta^\star
	\omega y}, \frac{500\nu}{y^2\omega}\right), \frac{4\rho \sigma_{\omega_2} k }{CD_{k\omega}y^2}\right)\right)^4\right),
$$ 
takes a value of 1 at the near wall region to activate the original equation for $\omega$ , and gradually switches to 0 moving away from the surface to activate the transformed $k-\epsilon$ equation. Here, $y$ denotes the distance to the nearest surface and 
$$
CD_{k\omega} = \max\left(\frac{2\rho\sigma_{\omega_2}}{\omega} \frac{\partial k}{\partial x_j}\frac{\partial \omega}{\partial x_j}, 10^{-10}\right).
$$
The turbulent eddy viscosity is defined as
\begin{equation}
	\nu_t = \frac{a_1k}{\max(a_1\omega, SF_2)},
\end{equation}
where $S$ denotes invariant measure of strain rate and $F_2$ is the second blending function that determines the value of $\nu_t$ to be taken and is given by 
$$
F_2 = \tanh\left(\left(\max \left(2\frac{\sqrt{k}}{\beta^\star
	\omega y}, \frac{500\nu}{y^2\omega}\right) \right)^2\right).
$$ 
The values of the constants are set as in \cite{menter2003ten}, $\beta^\star = 0.09$, $\alpha_1=5/9$, $\beta_1 = 3/40$, $\sigma_{k_1}=0.85$, $\sigma_{\omega_1}=0.5$, $\alpha_2=0.44$, $\beta_2=0.0828$, $\sigma_{k_2}=1$, $\sigma_{\omega_2}=0.856$.

\bibliography{references}
\bibliographystyle{ieeetr}

\end{document}